\documentclass[a4paper,11pt]{article}
\pdfoutput=1 % if your are submitting a pdflatex (i.e. if you have
\usepackage{jcappub} % for details on the use of the package, please
\usepackage[T1]{fontenc} % if needed

\newcommand{\beq}{\begin{equation}}
\newcommand{\eeq}{\end{equation}}
\newcommand{\beqa}{\begin{eqnarray}}
\newcommand{\eeqa}{\end{eqnarray}}
\newcommand{\beqar}{\begin{eqnarray*}}
\newcommand{\eeqar}{\end{eqnarray*}}

\begin{tiny}

\end{tiny} %{\vskip-2ex$_{#1}%$\label{#1}}

\title{Diffuse flux of
galactic neutrinos and gamma rays}

\vspace{50pt}
\author{J.M.~Carceller,} 
\author{M.~Masip}
\vspace{16pt}

\affiliation{CAFPE and Departamento de F{\'\i}sica Te\'orica y del Cosmos\\
Universidad de Granada, E-18071 Granada, Spain}
\vspace{16pt}

\emailAdd{jmcarcell@correo.ugr.es}
\emailAdd{masip@ugr.es}

\abstract{
We calculate the fluxes of neutrinos and gamma rays from
interactions of cosmic rays with interstellar matter 
in our galaxy.
We use EPOS-LHC, SIBYLL and GHEISHA to parametrize
the yield  of these particles in proton, helium and 
iron collisions at kinetic energies between 1 and $10^8$ GeV,
and we correlate the cosmic ray density with the mean 
magnetic field strength in the disk and the halo of our
galaxy. We find that at $E > 1$ PeV the
fluxes depend very 
strongly on the cosmic-ray composition, whereas at 
1--5 GeV the main source of uncertainty is the cosmic-ray 
spectrum out of the heliosphere. 
We show that the diffuse flux of galactic
neutrinos becomes larger than the conventional 
atmospheric one at $E>1$ PeV, but that  at all IceCube energies
it is 4 times 
smaller than the atmospheric flux from forward-charm 
decays. }

\begin{document}
\maketitle
\flushbottom

\section{Introduction}

Cosmic rays (CRs) are a puzzling component of our universe. 
The simplicity of their spectrum between $10$ and $10^{11}$ GeV, a
power law $E^{-\alpha}$ with $\alpha\approx 2.7$, is apparent but
very surprising, since it
suggests that nature has used the same strategy to accelerate
all of them. It also indicates that their propagation from the 
sources to the Earth is universal, in the sense that it scales 
trivially with the CR energy. Actually, 
both ends of the power law describing their spectrum 
could be mere propagation
effects caused by the heliosphere (acting
as a magnetic barrier 
at CR energies below 10 GeV) and
by the 2.7K CMB (causing inelastic collisions 
above $10^{11}$ GeV \cite{Greisen:1966jv,Hooper:2006tn}).
A closer look at the spectrum (see \cite{Blasi:2014ava} for 
a review), however, reveals features
that may have an important physical meaning:
at the $10^{6.5}$ GeV {\it knee} the spectral index
changes from 2.7 to 3, whereas at the $10^{9.5}$ GeV {\it ankle}
it goes back to 2.7. The knee and the ankle could signal, respectively, 
the energy limit of the main accelerators in our galaxy 
and the point where most CRs have an extragalactic origin.

Another remarkable feature in the CR flux is its almost 
perfect isotropy: TeV CRs reach the Earth 
with the same frequency 
(at the  0.1\% level \cite{Amenomori:2006bx}) 
from all directions. 
Being charged, CRs are coupled to the 
galactic and intergalactic magnetic fields, and their isotropy 
is obtained after very long trajectories that lose memory 
of the position of the sources. In particular, the 
time of flight of $10$--$100$ GeV CRs can be estimated
by comparing the ratio of light (Li, Be, B) to medium 
(C, N, O) nuclei in their composition \cite{Strong:1998pw}.
Light nuclei are much more abundant in CRs than in the solar system,
which reveals their origin as secondary particles produced in 
collisions of medium nuclei while crossing around 3 Mpc
of interstellar (IS) medium. This length, confirmed by 
the measurement of radioactive {\it clock} isotopes 
in the CR flux \cite{deNolfo:2006qj}, 
is much larger than the 0.3 kpc thickness or 
the 30 kpc diameter of our galaxy.
CRs behave like a very diffuse gas trapped in the galaxy by 
magnetic fields \cite{shalchi}. 

Up to energies close to the knee the main source of CRs could be
galactic supernova remnants. Simple arguments suggest
a spectral index at the acceleration site around $\alpha_0\approx 2$ 
independent from the charge or the mass of the CR \cite{gaisser}. 
The observed spectrum, with $\alpha=2.6$--$2.7$, would then result
after including propagation effects; as more energetic CRs
escape faster into the halo, their density in the galactic disk
will be reduced by a factor of $\approx E^{-0.5}$ (typical for 
a Kraichnan spectrum of magnetic turbulences \cite{shalchi}). 
Satellite and balloon experiments have found 
that the CR composition is dominated by hydrogen and $^4$He 
nuclei (at $E<E_{\rm knee}$ we can neglect the presence of 
heavier nuclei), 
and that both species have slightly 
different spectral indexes \cite{Boezio:2012rr,Yoon:2011aa}. 
Between 10 and $10^{6.7}$ GeV we estimate
\beq
\Phi_p = 1.3 \left( {E\over {\rm GeV}} \right)^{-2.7}
\; {\rm protons / ( cm^2\,s\,sr \, GeV)}
\label{fluxp}
\eeq
and 
\beq
\Phi_{\rm He} = 0.54 \left( {E\over {\rm GeV}} \right)^{-2.6}
\; {\rm nuclei / (cm^2\,s\,sr \, GeV)}\,.
\label{fluxHe}
\eeq
This implies an all-nucleon flux around
$\Phi_N \approx 1.8 \times 10^4 \,( {E/ {\rm GeV}} \,)^{-2.7}
\;{\rm ( m^2\,s\,sr \, GeV)^{-1}}$ and also that the number of
protons and He nuclei becomes similar at 
$E\approx 10$ TeV.
Notice that the number of helium to hydrogen nuclei in CRs 
is much larger than the 1 to 12 ratio 
obtained from Big-Bang nucleosynthesis.
Between the knee and the ankle ({\it i.e.},  
$10^{6.7}\; {\rm GeV}< E < 10^{9.5}\; {\rm GeV}$) the CR 
composition is uncertain, while the spectrum becomes
\beq
\Phi = 330 \left( {E\over {\rm GeV}} \right)^{-3.0}
\; {\rm nuclei / ( cm^2\,s\,sr \, GeV)}\,.
\label{fluxA}
\eeq

Information about CRs comes not only from direct 
measurements, it can be completed by observations in other
{\it channels}: CR interactions with IS matter 
produce a diffuse flux of secondary  neutrinos and gamma rays 
that is sensitive to their composition and spectrum,
and also to the gas
distribution in our galaxy. At TeV--PeV energies these galactic neutrinos 
could define a flux observable at 
IceCube \cite{Halzen:2010yj}, whereas lower energy gamma rays 
may bring information, for
example, 
about the CR density out of the heliosphere. At any rate, these
secondary 
particles are a certain background in the search for astrophysical
sources at neutrino and gamma-ray observatories.

In this paper we will attempt a complete calculation of the 
neutrino and gamma-ray fluxes from CR spallation 
inside our galaxy \cite{Stecker:1978ah,Berezinsky:1992wr,Strong:1998fr,Evoli:2007iy,Murase:2013rfa,Neronov:2013lza,Joshi:2013aua,Ahlers:2013xia,Kachelriess:2014oma,Dado:2015eda,Ahlers:2015moa},
paying special attention to the energy and composition
dependence of the yields. In particular, we will use
the MonteCarlo simulators EPOS-LHC \cite{Pierog:2013ria}, 
SIBYLL \cite{Ahn:2009wx} and (at $E<50$ GeV) 
GHEISHA \cite{Jakubowski:1989vy} to
obtain precise parametrizations of the 
$(\gamma,\nu_i,\bar \nu_i)$ yields for proton, helium and iron
primaries at different energies. We will also define a scheme
where the CR density inside our galaxy is
correlated with the mean magnetic field strength in the disk
and the halo. This and the gas distribution will provide 
the angular dependence of the gamma-ray and
neutrino fluxes both at IceCube 
and Fermi-LAT \cite{TheFermi-LAT:2015kwa} energies.

\section{Neutrinos and gammas from galactic CRs}
Consider a point $\vec r = r\, \vec u_r$ 
in our galaxy where the CR number density for the species 
$A = (p,\,{\rm He},...,\,{\rm Fe})$ 
is $n_{A}(E,\vec r)$ and where the IS gas density is 
$\rho_{i}(\vec r)$, with $i=(p,\,$He). 
If the CR distribution is isotropic, the number density and
the flux $\Phi_{A}(E,\vec r)$ at that point
are simply related:
\beq
\Phi_{A}(E,\vec r) = {c \over 4\pi}\; n_{A}(E,\vec r)\,.
\eeq
Neutrinos from collisions
of type $A$ CRs with type $i$ IS gas that reach
the Earth (at $\vec r=0$) from
direction $\vec u_r$ may be produced at any point along
the line generated by $\vec u_r$. The total neutrino
flux is then \cite{gaisser,Lipari:1993hd}
\beq
\Phi_\nu(E,\vec u_r) = \sum_{A,i}\, \int_0^\infty {\rm d} r \int_E^\infty
{\rm d} E'\;{1\over \lambda_{Ai}(E',\vec r)} \; \Phi_A(E',\vec r) \;
{{\rm d} n^{\nu}_{A i}\over {\rm d} E}(E;E')\,,
\label{flux}
\eeq
where the interaction length for $A\,i$ collisions
at CR energy $E'$ reads
\beq
\lambda_{Ai}(E',\vec r)={m_i\over \rho_{i}(\vec r)\;
\sigma_{A i}(E')}
\eeq
and ${{\rm d} n^{\nu}_{A i}/ {\rm d} E}$ expresses
the yield of neutrinos of energy $E$ obtained in those
collisions.
An analogous expression including 
the transparency $\eta(E,\vec r)$ of the galaxy would describe
the gamma-ray flux reaching the Earth. It is revealing  
to rewrite Eq.~(\ref{flux})
in terms of the fraction $x=E/E'$ of energy taken by the neutrino:
\beq
\Phi_\nu(E,\vec u_r) = \sum_{A,i}\, \int_0^\infty {\rm d} r \; 
\rho_{i}(\vec r) \int_0^1
{\rm d} x\;{\sigma_{A i}(E/x) \over m_{i}} \; 
\Phi_A(E/x,\vec r) \; x^{-1} 
f^{\nu}_{A i}(x,E/x)\,,
\label{flux-y}
\eeq
with $f^{\nu}_{A i}(x,E')$ the yield of neutrinos carrying a fraction
$x$ of the CR energy $E'$. This equation goes into a much simpler
form after several assumptions that are common in the literature
(in next sections we will check their validity):
\begin{enumerate}
\item Since neutrinos come
from pions and these are created in nucleon-nucleon collisions,
one expects that the yield from a He nucleus of energy $E$ coincides
with that from a proton of energy $E/4$, {\it i.e.}, the yield is 
a function of the energy per nucleon in the CR. This motivates the 
definition of a flux $\tilde \Phi_{CR}(E)$ of CRs per unit of energy 
per nucleon,\footnote{Notice that $\tilde \Phi_{CR}$ is {\it not} 
the all-nucleon flux $\Phi_{N}(E)=\sum_A A^2 \,\Phi_{A}(A\,E)$ used
to calculate atmospheric lepton fluxes.}
\beq
\tilde \Phi_{CR}(E,\vec r)=\sum_A\tilde \Phi_A(E,\vec r)=
\sum_A A \,\Phi_{A}(A\,E,\vec r)\,.
\label{fluxperN}
\eeq
\item The spectrum of secondaries after the fragmentation
of a high-energy CR does not depend on the target (the H or He nucleus 
in the IS gas). 
\item One may take a constant spectral index $\alpha$ for the 
CR flux in the energy interval of interest.
\item The CR density is 
homogeneous inside the galactic disk, like in the usual 
leaky-box model:
$\tilde \Phi_{CR}(E,\vec r)= C\, E^{-\alpha}\,$. Later on 
we will correlate the CR flux
with the magnetic fields in the halo and the disk.
\item For the cross sections
fixing the interaction lengths, one may neglect their
energy dependence and assume a $A^{2/3}$ scaling
with the mass number of the
target and the projectile. 
\end{enumerate}
After all these approximations,
the sum over 
the different species $A$ in the CR flux 
and over the two components in the IS gas (around 3 grams
of hydrogen per each gram of helium) gives 
\beqa
\Phi_\nu(E,\vec u_r) &=& C\, E^{-\alpha} \,f
\,{\sigma_{p p} \over m_{p}} \,
\int_0^\infty {\rm d} r \; \rho_{IS}(\vec r) \,
\int_0^1 {\rm d} x\;
 x^{\alpha -1} 
f^{\nu}_{p p}(x) \cr
&=& {C\, f\, \sigma_{p p} \over m_{p}} \,
X(\vec u_r) \, Z_{\nu p}^{(\alpha-1)}\, E^{-\alpha} \,,
\label{fluxapp}
\eeqa
where $\rho_{IS}(\vec r)$ is the total mass density at $\vec r$,
$X(\vec u_r)$ is the total depth of IS matter along the direction $\vec u_r$,
and $Z_{\nu p}^{(\alpha-1)}$ is the ($\alpha -1$)-moment  of the
neutrino yield in $pp$ collisions. The parameter 
\beq
f = \left( {3 + 4^{-1/ 3} \over 4} \right)
\;\sum_A r_A A^{2\over 3} \,,
\label{f}
\eeq
with $r_A$ the fraction of the primary $A$ in the total flux 
given by Eq.~(\ref{fluxperN}) ($r_p\approx 0.9$ and 
$r_{\rm He}\approx 0.1$ for the fluxes in 
Eqs.~(\ref{fluxp},\ref{fluxHe})), takes into account
the mixed composition both of the IS gas and
of the CR flux.

The expression in Eq.~(\ref{fluxapp}) implies neutrino 
and gamma-ray fluxes with 
the same spectral index as the primary CRs and
proportional both to the total depth (column 
density) along each direction in our galaxy and to the order-1.7
$Z$-moment of the yield in $pp$ collisions. 
Let us use MonteCarlo simulators to analize
in detail cross sections and yields and establish the validity of 
these approximations at different energies.

\section{Neutrino and gamma-ray yields}

\begin{figure}[!t]
\begin{center}
\includegraphics[width=0.48\linewidth]{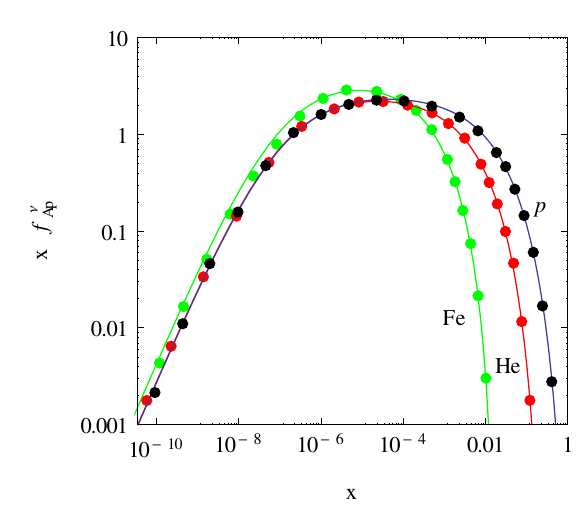}\hspace{0.3cm}
\includegraphics[width=0.48\linewidth]{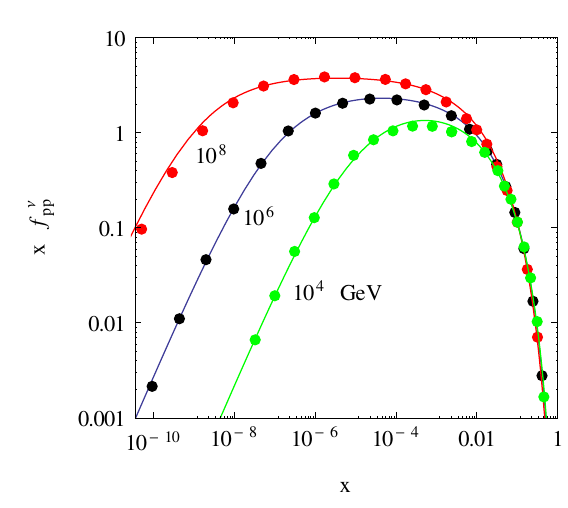}
\end{center}
\caption{{\bf Left.} $\nu_e$ yield $f^{\nu_e}_{Ap}(x,10^6\,{\rm GeV})$ from
$A=$ proton, helium and iron collisions
obtained with EPOS-LHC (dots) and our 2-parameter fits (solid).
{\bf Right.} $f^{\nu_e}_{pp}(x,E)$ for $E=10^4$ GeV,
$10^6$ GeV and $10^8$ GeV. 
\label{fig1}}
\end{figure}
We plot in Fig.~\ref{fig1}--left the $\nu_e$ yield  from the
collision of a $10^{6}$ GeV proton ($A=1$), He ($A=4$) or
Fe ($A=56$) with a proton at rest. 
We have obtained these average yields after the simulation
of $10^4$ events with EPOS-LHC. In the plot we include 
a 2-parameter fit (solid line) of type
\beq
f^{\nu}_{Ap}(x,E)= a \,{\left(1 - A x\right)^3\over x}\,
{e^{-b\,(A\, x)^{0.43}}\over \left( 1 + \displaystyle 
\sqrt{0.1 \; {\rm GeV}\over xE}\, \right)^{\!2} }\,,
\label{fit1}
\eeq
with
$a(E,A)$ and $b(E,A)$ the energy and mass-number 
dependent parameters. 
In Fig~\ref{fig1}--{right} we explore how these yields change
with the energy.
For the gamma yield we use the same type of fit changing 
$0.1\;{\rm GeV} \to 0.2\;{\rm GeV}$. 
We find that at high 
energies Eq.~(\ref{fit1}) provides surprisingly precise
fits for all the yields in proton, helium and even 
iron collisions. The two parameters in the fit
change logarithmically with the energy according to 
\beqa
a&=&a_0  \left(1+0.073\, \ln {E\over 10^6\,{\rm GeV}} +
0.0070\, \ln^2 {E\over 10^6\,{\rm GeV}} \right) \\
b&=&b_0  \left( 1+ 0.020\, \ln {E\over 10^6\,{\rm GeV}} +
0.0018 \,\ln^2 {E\over 10^6\,{\rm GeV}} \right) \,.
\eeqa
In Table 1--left we list the numerical 
values of $a_0$ and $b_0$ that fit the yield of 
(anti)neutrinos of each flavor in $pp$ collisions, whereas
in Table 1--right we provide our best fit to the total neutrino
and gamma yields for different CR primaries. 
\begin{table}
\begin{center}
\begin{tabular}{r|c|c|c|c|}
          \multicolumn{1}{c}{}
        & \multicolumn{1}{c}{$\nu_e$} 
        & \multicolumn{1}{c}{$\bar \nu_e$} 
        & \multicolumn{1}{c}{$\nu_\mu$} 
        & \multicolumn{1}{c}{$\bar \nu_\mu$} \\
\multicolumn{5}{c}{} \\ [-3ex]
\cline{2-5}
$a_0$  &  2.7  & 2.7 & 5.1 & 5.1  \\
\cline{2-5}
$b_0$  & 7.7  & 8.7 & 8.3 & 8.3  \\
\cline{2-5}
\multicolumn{5}{c}{}\\[-2.5ex]
\multicolumn{1}{c}{}
&\multicolumn{4}{c}{$p\,p$} 
\end{tabular}
\hspace{0.5cm}
\begin{tabular}{r|c|c||c|c||c|c|}
          \multicolumn{1}{c}{}
        & \multicolumn{1}{c}{$\nu$} 
        & \multicolumn{1}{c}{$\gamma$} 
        & \multicolumn{1}{c}{$\nu$} 
        & \multicolumn{1}{c}{$\gamma$} 
        & \multicolumn{1}{c}{$\nu$} 
        & \multicolumn{1}{c}{$\gamma$} \\
\multicolumn{7}{c}{} \\ [-3ex]
\cline{2-7}
$a_0$  &  15.5  & 5.8 & 14.8 & 5.9 & 20.0 & 8.8   \\
\cline{2-7}
$b_0$  & 8.0  & 5.3 & 7.3 & 5.0 & 5.8 & 4.0 \\
\cline{2-7}
\multicolumn{6}{c}{}\\ [-2.5ex]
\multicolumn{1}{c}{}
&\multicolumn{2}{c}{$p\,p$} 
&\multicolumn{2}{c}{${\rm He}\,p$} 
&\multicolumn{2}{c}{${\rm Fe}\,p$} 
\end{tabular}
\end{center}
\vspace{-0.5cm}
\caption{Parameters $a_0$ and $b_0$ obtained from the 
EPOS-LHC simulation for the 
yield of each neutrino species in $pp$ collisions (left)
and for the 
total yield of neutrinos and gammas in proton, He and Fe 
collisions with a proton at rest (right). 
\label{par1}}
\end{table}
Analogous parametrizations have been obtained, for example, 
by Kamae et al.~\cite{Kamae:2006bf}, who separate diffractive
and non-diffractive contributions, 
or by Kelner and collaborators \cite{Kelner:2006tc}. 
Our EPOS-LHC simulations
include neutrinos and gammas from any unstable particle produced 
(directly or through a complete set hadronic resonances) in these
collisions: not only the leading contribution from pions 
discussed in these references but also from kaons, as well as
etas, omegas, neutrons, etc. This may slightly change the spectrum
or, for example, also the ratio of flavors in the neutrino yields. 
In addition, our two parameter fit 
works well also in the collisions of heavy nuclei 
(it is straightforward to
interpolate the values of $a$ and $b$ for any nucleus
between helium and iron).

Repeating
the simulations with SIBYLL we obtain very similar results. For 
example, at $10^6$ GeV we obtain a multiplicity 
(the 0-moment of the yield) of 44 $\nu_e+\bar \nu_e$, 
83 $\nu_\mu+\bar \nu_\mu$ and 47 gammas with EPOS-LHC
versus 41, 83 and 50 with SIBYLL, respectively. 
For the 1.7-moment, we obtain identical
results for $Z_{\nu_\mu p}^{(1.7)}+ Z_{\bar \nu_\mu p}^{(1.7)}$ 
but 9\% and 7\% larger values with EPOS-LHC for 
$Z_{\nu_e p}^{(1.7)}+ Z_{\bar \nu_e p}^{(1.7)}$ and 
$Z_{\gamma p}^{(1.7)}$, respectively. In our calculation of the galactic fluxes
we will take the yields from EPOS-LHC and use SIBYLL to estimate 
the uncertainty. These hadronic simulators do not provide distributions
at $E<50$ GeV; we have checked, however, that the yields
obtained with GHEISHA agree reasonably well (the differences in 
the $Z$-moments are within a 15\%) with the values provided
by the fit in
the 1--50 GeV interval of energy.
Therefore, we will use our fit 
in the whole $1$--$10^8$ GeV interval of kinetic 
energies.

As for the different approximations discussed in the
previous section,
\begin{itemize}
\item The dependence of the yields on the (H or He) target that 
fragments the CR is small. We find
that $Z_{\nu p}^{(1.7)}$ is just a factor of $c_T\approx 1.02$ larger
when the target is a He nucleus instead of a proton. The presence of 1 g
of He per 3 g of H in the IS medium can then be accounted for by  
the factor 
\beq
F_{A} = \frac{3}{4} + \frac{c_T}{16}\, 
\frac{\sigma_{A {\rm He}}}{\sigma_{A p}}\,,
\eeq 
that corrects the  $(3 + 4^{-1/ 3})/ 4$ factor 
in Eq.~(\ref{f}). For the
three CR primaries that we will consider we
obtain 
$F_p\approx 0.92$, $F_{\rm He}\approx 0.90$ and $F_{\rm Fe}\approx 0.86$.
\item At CR energies above 1 TeV the $A\,p$ cross section is not
constant but a power law:
\beq
\sigma_{A p}(E) = \sigma^0_{A p}\, 
\left( {E\over 1\; {\rm GeV}} \right)^{\beta_{A}} \,.
\eeq
In $p\,p$ collisions using EPOS-LHC we obtain $\beta_p=0.082$ and 
$\sigma^0_{p p}=17.7$ mb, whereas in 
He$\,p$ and Fe$\,p$ collisions we find
$\beta_{\rm He}=0.062$, $\sigma^0_{{\rm He}\, p}=60.5$ mb, 
$\beta_{\rm Fe}=0.026$ and $\sigma^0_{{\rm Fe}\, p}=551$ mb.
This energy dependence will 
change the spectral index of the galactic neutrinos from  
$\alpha_A$ to $\alpha_A-\beta_A$. For this reason, 
in our calculation we will separate
the contributions of each component in the CR flux 
({\it i.e.}, we will not use 
the approximations 1, 3 and 5 discussed in the previous section).
\item We will assume a (position-dependent) CR flux of type
$\Phi_{A}(E,\vec r)= g(\vec r)\,\Phi^0_{A}(E)\,$, 
where $g(\vec r)$ is the ratio between the flux at 
position $\vec r$ in
our galaxy 
and the flux $\Phi^0_{A}(E)$ that we observe at the Earth 
($r=0$). Eq.~(\ref{flux-y}) becomes then
\beq
\Phi_\nu(E,\vec u_r) = \sum_A \frac{F_A}{m_p} 
\int_0^\infty 
{\rm d}r \, g(\vec r)\, \rho_{IS}(\vec r) 
\int^1_0 {\rm d}x \, \sigma_{A p}(E/ x)\, 
\Phi^0_A(E/x)\, x^{-1} f^{\nu}_{A}(x,E/x) \,.
\label{final1}
\eeq
This is the final equation that we will 
use to find the neutrino and the gamma-ray fluxes. The
sum runs over the different species $A$
in the CR flux, while the effect of the He fraction present in the 
IS matter is encapsulated in the coefficients $F_A$. 
The energy dependence in this equation simplifies
if one assumes
that both the CR fluxes 
and the cross sections follow unbroken 
power laws and that the yields are energy independent:
\beq
\Phi_\nu(E,\vec u_r) = \sum_A \frac{F_A \, \sigma^0_{A p}\, C_A }{m_p} \;
E^{-(\alpha_A-\beta_{A})} \int_0^\infty 
{\rm d}r \, g(\vec r)\; \rho_{IS}(\vec r)\, 
\int^1_0 {\rm d}x\, x^{\alpha_A-\beta_{A}-1} \,
f^{\nu}_{A}(x) \,,
\label{final0}
\eeq
where $\Phi^0_{A}(E)=C_A E^{-\alpha_A}$ is the flux of CR
nuclei type $A$ at the Earth. 
We will compare the results provided by this 
expression with the more general ones from 
Eq.(\ref{final1}).
\end{itemize}

\section{Angular dependence of the fluxes}
The integral 
\beq
\tilde X_\nu(\vec u_r) \equiv \int_0^\infty \: 
{\rm d}r \, g(\vec r)\; \rho_{IS}(\vec r)
\label{ang}
\eeq
that appears in Eq.~(\ref{final1}) provides all the angular dependence
of the galactic neutrino fluxes. Basically, $\tilde X_\nu(\vec u_r)$ is the 
depth of IS matter along the direction $\vec u_r$ corrected by the
CR density at each point relative to the one that we observe at 
the Earth. The analogous expression for high-energy gamma rays will include 
an attenuation factor $\eta(E,\vec r)$ from gamma-ray 
collisions with the CMB 
or with infrared galactic light \cite{aharonian}:
\beq
\tilde X_\gamma(E,\vec u_r) \equiv \int_0^\infty \: 
{\rm d}r \, g(\vec r)\; \eta(E,\vec r)\; \rho_{IS}(\vec r)\,.
\label{ang-gamma}
\eeq
In particular,
the mean free path for the conversion of a $E=0.5$--$5$ PeV 
gamma ray into an $e^+e^-$ pair through the collision with a CMB
photon is around 10 kpc, whereas if the gamma travels 
near the galactic center (at radial distances below 4 kpc) 
it may experience the same type of conversion at 
energies 10 times smaller.

For the IS matter in our galaxy, we will take a total mass 
of $4.6\times 10^{11}\,M_\odot$, 
which is $13\%$ of all its baryonic mass. We will distribute 
this mass in
the thin disk, the thick disk and the galactic halo according the  
basic scheme described in \cite{kalberla}. The thin disk reaches up to
$R\approx 35$ kpc, with a radial scale $R_n=3.15$ kpc and a
vertical scale height $h_t$ that increases with $R$:
\beq
\rho_{t}(R,z) =
\left\{\begin{array}{ll}
m_p\,n_{t}\; \displaystyle \exp{\left(-\frac{R-R_\odot}{R_n} 
-\frac{z\ln 2 }{h_{t}(R)}\right)} &  R > 7\;{\rm kpc}\\
\\
m_p\,n_{t}\; \displaystyle \exp{\left(-\frac{7-R_\odot}{R_n} 
-\frac{z\ln 2 }{h_{t}(R)}\right)} &  R < 7\;{\rm kpc}
\,,
\end{array}\right.
\eeq
where $n_t=1.5$ cm$^{-3}$, $R_\odot=8.4$ kpc, 
$h_t(R) = h_{t0}\, e^{\frac{R-R_0}{R_0}}$, 
$h_{t0}=0.15$ kpc and $R_0=9.8$ kpc. The thick disk has a similar 
structure but a different density and scale height:
$n_T=0.15$ cm$^{-3}$ and $h_{T0}=0.4$ kpc, respectively. Finally, we will take a
60 kpc halo with spherical symmetry and a large radial scale:
\beq
\rho_{h}(R,z) = m_p\,n_{h}\; 
\exp{\left(-\frac{\sqrt{R^2+z^2}-R_\odot}{r_h}\right)}\,,
\eeq
with $n_h=0.001$ cm$^{-3}$ and $r_h=12$ kpc.

For the galactic CR density, we will modify the simple 
but effective leaky-box model, that assumes a constant density 
in the galactic disk, by correlating it 
with the mean magnetic field strength 
at each point in the disk and the halo. 
As we have argued in the introduction, the CR spectral index 
$\alpha_0=2.0$--$2.2$ 
at the sources is {\it changed} into the one we see, 
$\alpha=2.6$--$2.7$, by the propagation through the regular and
turbulent magnetic fields in the IS medium. The extra 
suppression of $\propto E^{-\beta}$ with $\beta\approx 0.5$
would reflect that higher 
energies have a larger diffusion coefficient 
and a shorter time of escape from
our galaxy (a position-dependent coefficient is used, for 
example, by the CR propagation code DRAGON \cite{Evoli:2008dv}). 
Since all magnetic effects depend on 
$R/B\approx E/(ZeB)$,
the flux of a very diffused CR gas should scale 
\beq
{\Phi_A(E,\vec r)\over \Phi_A(E, 0)} = g(\vec r) \approx 
\left( {B(\vec r)\over B_0} \right)^{0.5}\,
\label{magnetic}
\eeq
independently of the position of the sources.
\begin{figure}[!t]
\begin{center}
\includegraphics[width=0.85\linewidth]{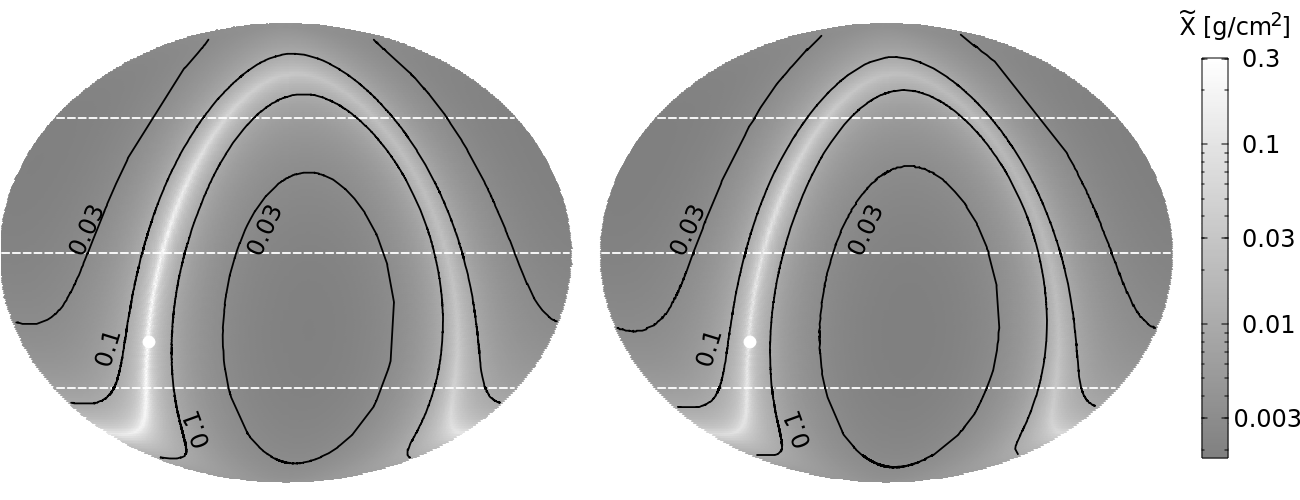}
\end{center}
\caption{{\bf Left.} Effective depth along the different directions
(equatorial coordinates). We indicate with
dashes the declinations $\delta= 0^\circ,\pm 45^\circ$ 
and with a white dot the galactic center. 
{\bf Right.} Effective depth for 0.5--5 PeV photons (it includes the
attenuation by the CMB).
\label{fig2}}
\end{figure}
Eq.~(\ref{magnetic}) expresses that CRs stay longer in regions
of the galaxy with stronger magnetic fields and, as a consequence,
their density in those regions is larger.
We will assume that in the disk the magnetic field
decays exponentially
with a radial scale $R_B=8.5$ kpc \cite{Han:2006ci}:
$B(R)=B_0\, e^{-(R-R_{\odot})/R_B}$ with $B_0=4\;\mu$G, and that 
in the halo it has a constant value around $B_h=1\;\mu$G.
Our hypothesis implies then that 
the CR density in the galactic center is
1.6 times larger than the one
we observe here, whereas in the halo it is 
a factor of 0.5 smaller.

\begin{figure}[!t]
\begin{center}
\includegraphics[width=0.49\linewidth]{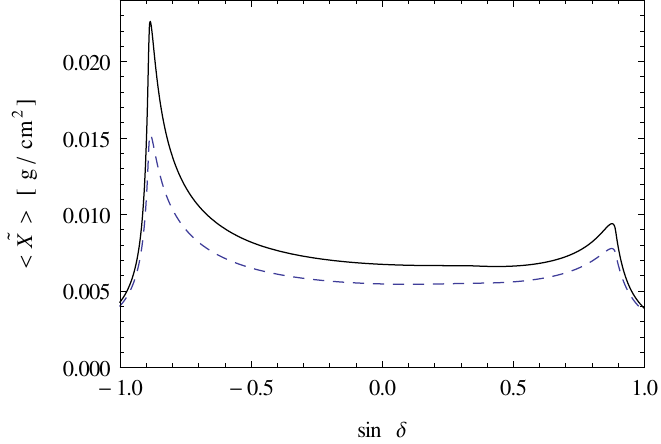}\hspace{0.3cm}
\includegraphics[width=0.47\linewidth]{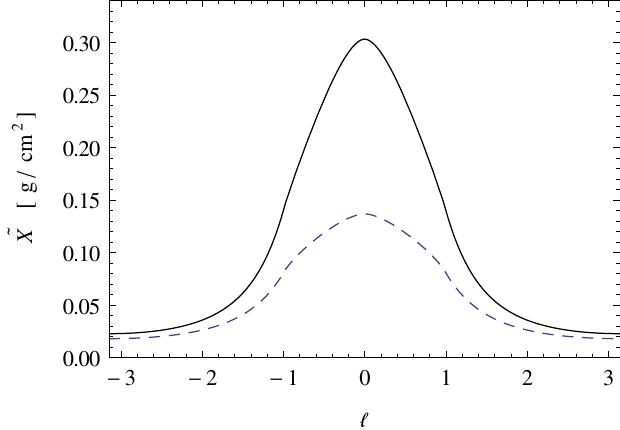}
\end{center}
\caption{{\bf Left.} Effective depth at different declinations 
averaged over the R.A. for neutrinos (solid)
and for 0.5--5 PeV gamma rays (dashes). {\bf Right.} Effective 
depth of the galactic disk (latitude $b=0$) at different
longitudes $\ell$.
\label{fig3}}
\end{figure}

Fig.~\ref{fig2} summarizes our results for the angular distribution
of the fluxes.
The plot on the left applies to neutrinos of any energy and
gamma rays with $E_\gamma < 50$ TeV. It gives a largest  
effective depth of 0.302 g/cm$^2$ 
from the galactic center at ${\rm R.A.} = -93^\circ$ and 
$\delta = -29^\circ$ and just 0.0019 g/cm$^2$ 
from the {\it darkest} point, at ${\rm R.A.} = -170^\circ$ and 
$\delta = 29^\circ$. The average depth is 
$\langle \tilde X_\nu\rangle = 0.0080$ g/cm$^2$.
The plot on the right refers to photons of 
0.5--5 PeV, which experience a maximum attenuation by the CMB. 
$\tilde X_\gamma$ varies between 
0.137 g/cm$^2$ and 0.0018 g/cm$^2$, with an average value of 
0.0064 g/cm$^2$.
In Fig.~\ref{fig3}--left we plot the average depth for 
each declination ({\it i.e.}, we have 
integrated over the R.A. and divided by $2\pi$),
whereas Fig.~\ref{fig3}--right gives the depth of the galactic
disc at different longitudes $\ell$.
We deduce, for example, that the ratio of 
positive versus negative declinations 
is approximately 3:4.
Neglecting the attenuation  of the 
neutrino flux by the Earth, this will also be the approximate 
ratio of {\it upgoing} 
versus {\it downgoing} neutrinos of galactic origin reaching IceCube. 

\section{Neutrino and gamma-ray fluxes}
We can finally give the neutrino and
gamma-ray fluxes from CR collisions with IS matter. 
At energies below the knee we have separated the proton
and the helium contributions (see the fluxes in 
Eqs.~(\ref{fluxp},\ref{fluxHe})). Since at $E\le 10$ GeV 
the effects of the
heliosphere on the CR flux are important,
in the 1--10 GeV interval we have considered 
two possibilities:
an unbroken power law ({\it i.e.},
the same spectral index $\alpha$ 
that holds at higher energies) or its reduction
in one unit (which gives the approximate CR 
flux that we see at low energies).
At $E>E_{\rm knee}$ we have studied different
CR compositions: proton, He and Fe. A mixed composition can
be easily deduced by combining these three possibilities.

\begin{figure}[!t]
\begin{center}
\includegraphics[width=0.49\linewidth]{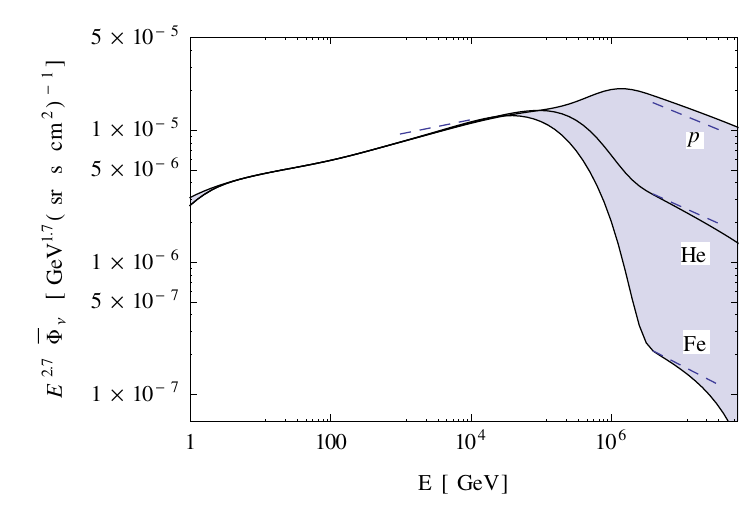}\hspace{0.4cm}
\includegraphics[width=0.46\linewidth]{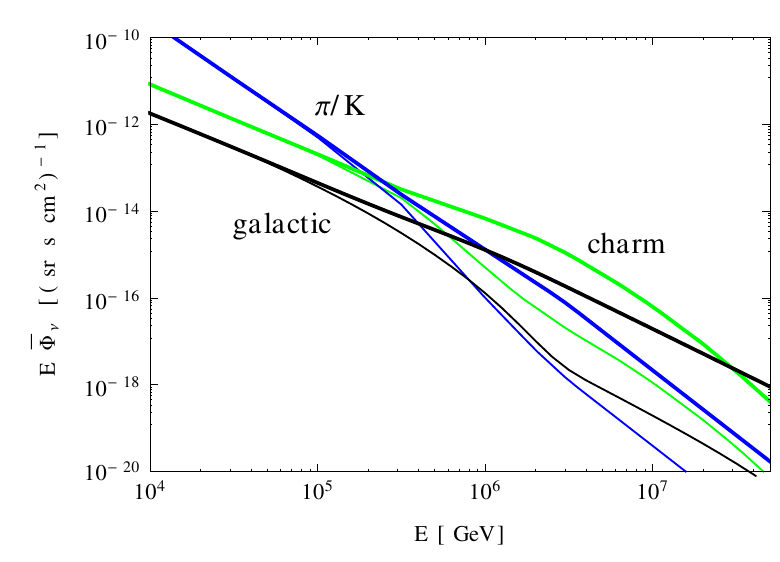}
\end{center}
\caption{{\bf Left.} Average (all directions) 
$\nu$ flux reaching the Earth. The shaded region
corresponds to different CR compositions ($p$, He
or Fe) at $E>E_{\rm knee}$. 
{\bf Right.} Comparison with the atmospheric $\nu$ fluxes from
pion and kaon decays and from forward charm
decays. For each contribution, we plot the $\nu$ flux for a 
proton (thick line) or an iron (thin line)  CR composition at $E>E_{\rm knee}$.
\label{fig4}}
\end{figure}
In Fig.~\ref{fig4}--left we give the neutrino flux ($\nu$ plus
$\bar \nu$ of the three flavors) obtained from Eq.~(\ref{final1}).
It is the flux averaged over all the directions
that reach the Earth. The shaded regions reflect the uncertainty
in the low-energy CR flux  and in the 
CR composition at $E>E_{\rm knee}$. In dashes we provide
the result derived from the expression in Eq.~(\ref{final0}).
Below $E_{\rm knee}$  this approximate result is 
[in ${\rm (cm^2\,s\, sr\,GeV)^{-1}}$]
\beq
\bar \Phi_\nu = 
3.7\times 10^{-6} \left({E\over {\rm GeV}}\right)^{-2.617}
+
0.9\times 10^{-6} \left({E\over {\rm GeV}}\right)^{-2.538}
\,,
\eeq
where the first and second terms correspond to the proton and the 
He contributions, respectively. The different spectral indexs 
are related to the different energy dependence of their
cross sections.
At energies above $E_{\rm knee}$ Eq.~(\ref{final0}) implies 
fluxes 
\beq
\bar \Phi_\nu = \left\{\begin{array}{ll}
4.4 \times 10^{-4} \, \left( {E\over {\rm GeV}}\right)^{-2.918} 
& {\rm \; (proton)}\,,\\ 
1.2 \times 10^{-4} \, \left( {E\over {\rm GeV}}\right)^{-2.938} 
& {\rm \; (helium)}\,,\\ 
1.3 \times 10^{-5} \, \left( {E\over {\rm GeV}}\right)^{-2.974} 
& {\rm \; (iron)}\,.
\end{array}\right.
\eeq
In Fig.~\ref{fig4}--right we compare this galactic flux with the 
atmospheric one. We see that the
conventional neutrino flux from light meson decays is larger than
the galactic one for all energies below 1 PeV. At 
$E\approx 250$ TeV the neutrino flux from forward charm decays
\cite{Halzen:2016thi,Gaisser:2013ira}
would dominate both the conventional
and the galactic fluxes. Notice that the atmospheric $\nu$ 
flux also has a strong dependence with the CR composition: it scales
like the all-nucleon flux, {\it i.e.}, like $1/A$ at $E>E_{\rm knee}$.

We may as well compare our results with the ones obtained by
other authors. For 
example, the galactic diffuse flux at 10--100 TeV deduced
with a GALPROP simulation (model $^SS^Z4^R20\,^T150\,^C5$) in
\cite{Ahlers:2015moa} is a 25\% smaller than the one we have 
found here, a difference that is within the uncertainty in
our (or any) calculation.

The relative frequency of the different neutrino and
antineutrino flavors in this flux is proportional
to their $Z$-moment of order 
$\alpha_A-\beta_A-1\approx 1.7$. For proton primaries, 
before $\nu$ oscillations we find 
\beq
(\nu_e : \nu_\mu : \nu_\tau : \bar \nu_e : \bar \nu_\mu : \bar \nu_\tau )
= {1\over 6}\, ( 1.29 : 1.89 : 0 : 0.87 : 1.95 : 0 )\,, 
\eeq
whereas in helium or iron collisions the relative $\nu$ and 
$\bar\nu$ abundances are closer:
\beq
(\nu_e : \nu_\mu : \nu_\tau : \bar \nu_e : \bar \nu_\mu : \bar \nu_\tau )
= {1\over 6}\, ( 1.11 : 1.96 : 0 : 1.06 : 1.87 : 0 )\,.
\eeq
This gives, in both cases, a 1.08:1.92 electron to muon ratio, 
which is a $10\%$
larger than the 1:2 ratio usually assumed from pion decays.
The distances that these galactic neutrinos  travel to reach the Earth
are much larger than their typical oscillation length. 
Taking $\theta_{12}=33^\circ$, $\theta_{23}=49^\circ$,
$\theta_{13}=8^\circ$ and $\delta_{\rm CP}=0$ 
\cite{Gonzalez-Garcia:2014bfa},
the probability for a flavor transition is then 
\beq
P_{\alpha \beta} = \sum_{j} |U_{\alpha j}|^2\, |U_{\beta j}|^2=
\left( \begin{array}{ccc}
0.549 & 0.222 & 0.229 \\ 
0.222 & 0.412 & 0.366 \\ 
0.229 & 0.366 & 0.405 
\end{array} \right)\,,
\eeq
with $U_{\alpha j}$ the PMNS matrix. For proton and He/Fe primaries, 
respectively, 
the final frequency of each flavor after oscillations is
\beq
(\nu_e : \nu_\mu : \nu_\tau : \bar \nu_e : \bar \nu_\mu : \bar \nu_\tau )
= {1\over 6}\, ( 1.13 : 1.07 : 0.99 : 0.91 : 0.99 : 0.91 )\,
\eeq
and
\beq
(\nu_e : \nu_\mu : \nu_\tau : \bar \nu_e : \bar \nu_\mu : \bar \nu_\tau )
= {1\over 6}\, ( 1.04 : 1.06 : 0.97 : 1.00 : 1.00 : 0.93 )\,. 
\eeq
\begin{figure}[!t]
\begin{center}
\includegraphics[width=0.47\linewidth]{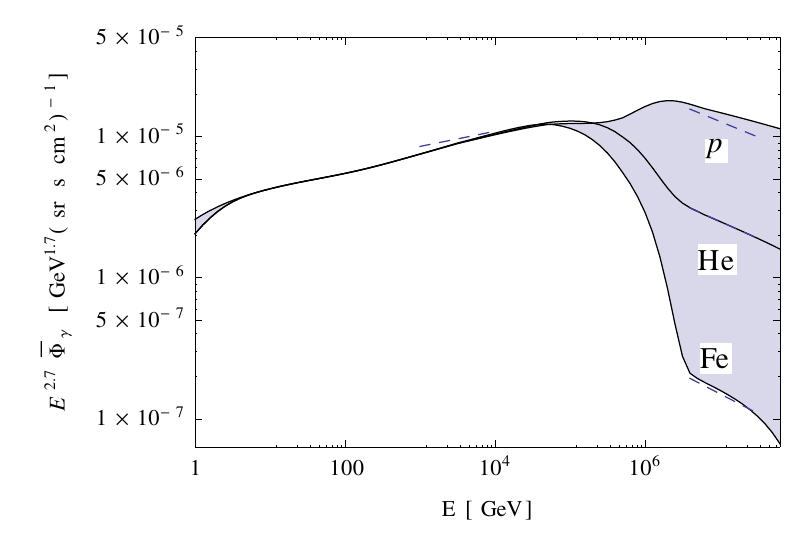}\hspace{0.4cm}
\includegraphics[width=0.48\linewidth]{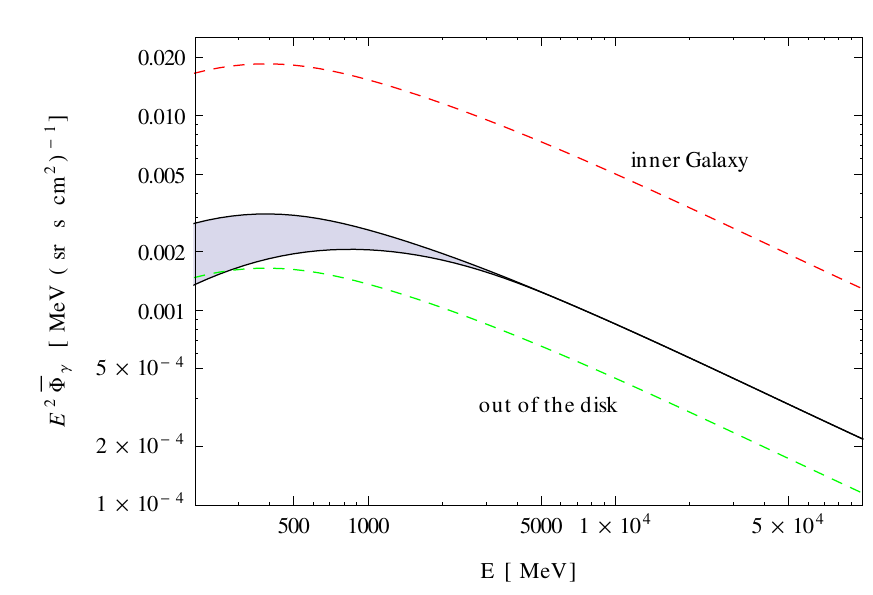}
\end{center}
\caption{{\bf Left.} Average gamma-ray flux from hadron fragmentation
(it does not include {\it leptonic} gamma rays). 
We plot with dashes
the approximate power laws deduced with the $Z$-moment method.
{\bf Right.} Average gamma-ray flux at low energies. We provide
the flux in from the inner Galaxy region 
($-8^\circ\le b\le +8^\circ$,  
$100^\circ\le \ell \le 260^\circ$) and from
outside the galactic disk ($|b|\ge 8^\circ$,  
$0^\circ\le \ell \le 360^\circ$) for an unbroken power law in 
the low-energy CR flux.
\label{fig5}}
\end{figure}

As for the  gamma-ray flux, 
our results are summarized in Fig.~\ref{fig5}. 
The approximate 
expression deduced
with the $Z$-moment method 
at energies below the knee  is
[in ${\rm (cm^2\, s\, sr\,GeV)^{-1}}$]
\beq
\bar \Phi_\gamma = 
3.4\times 10^{-6} \left({E\over {\rm GeV}}\right)^{-2.617}
+
0.8\times 10^{-6} \left({E\over {\rm GeV}}\right)^{-2.538}
\,,
\eeq
that includes a 0.93 attenuation factor 
from collisions with infrared light 
in the center of the galaxy \cite{Moskalenko:2005ng}.
At higher energies gamma-ray 
interactions with the CMB increase the suppression to 0.80, implying
\beq
\bar \Phi_\gamma = \left\{\begin{array}{ll}
4.2 \times 10^{-4} \, \left( {E\over {\rm GeV}}\right)^{-2.918} 
& {\rm \; (proton)}\,,\\ 
1.1 \times 10^{-4} \, \left( {E\over {\rm GeV}}\right)^{-2.938} 
& {\rm \; (helium)}\,,\\ 
1.2 \times 10^{-5} \, \left( {E\over {\rm GeV}}\right)^{-2.974} 
& {\rm \; (iron)}\,.
\end{array}\right.
\eeq
In Fig.~\ref{fig5}-right we give the low-energy flux
averaged over different galactic regions. At these energies the
gamma-ray flux observed by Fermi-LAT \cite{Ackermann:2012pya} 
includes several 
components, being the diffuse flux from galactic CR interactions 
obtained here the dominant one. 
We may compare our result with their estimate
using a GALPROP simulation; for example, 
in the inner galactic region ($-8^\circ\le b\le +8^\circ$ and 
$100^\circ\le \ell \le 260^\circ$) at $E\le 1$ GeV our 
gamma-ray flux is very similar to the
one they obtain \cite{Ackermann:2012pya} with GALPROP 
(model $^SS^Z4^R20\,^T150\,^C5$), but at 
$E=100$ GeV ours is 20\% larger. These differences are within
the expected uncertainties; incidentally, at 10-100 GeV
Fermi-LAT predicts a gamma-ray flux from the inner galactic
region that is below the one they observe.

\section{Summary and discussion}
Charged cosmic rays are the only known source of high-energy 
neutrinos and gamma rays. In addition to the Earth's atmosphere, 
in our galaxy one may distinguish two
different environments where these neutral particles are produced: 
the dense regions where CRs are accelerated (point-like sources), 
and the IS space
where CRs are trapped for several million years (diffuse flux). 
A precise 
characterization of the diffuse flux of galactic neutrinos
and gamma rays from CR
collisions with IS matter is necessary for the identification
of astrophysical sources \cite{Kistler:2006hp} or in the
search for other possible sources, like the annihilation
of dark-matter particles.

Here we have attempted a calculation of these fluxes that covers
the whole energy spectrum of interest. We have paid special 
attention to the yields of neutrinos and gamma rays
from CR fragmentation at different
energies. In particular, we provide a two-parameter fit that
gives an accurate description (within the 15\%) for the 
moments of order 0 (total number of particles), 1 (fraction
of energy taken by these particles) and 2 (dictates the
total high-energy fluxes) of the neutrino and photon distributions
obtained with
different simulators (the results with EPOS-LHC, SIBYLL or GHEISHA 
may differ in a $15\%$). 

In our study we have assumed a particular model for the distribution
of IS matter (hydrogen and helium) in the galaxy. 
Although this distribution may have a certain degree of uncertainty,
it reproduces well the total mass of the galactic gas, which is
more constrained and fixes the average neutrino
and gamma-ray fluxes in Figs.~\ref{fig4} and \ref{fig5}. We have also
assumed a galactic CR density correlated with the mean 
magnetic field strength in the disk and the halo. This 
should provide a
better description than the constant CR density in the
disk with an empty halo 
used in the usual leaky-box model. 

At any rate, we have shown that the two main sources 
of uncertainty in the galactic neutrino and gamma-ray 
fluxes are the low-energy CR spectrum out of the
heliosphere and the CR composition at energies 
beyond $E_{\rm knee}$. The first one gives 
$30\%$ variations in the gamma-ray flux at energies around
1 GeV, where Fermi-LAT has observed 
an excess from the galactic center \cite{Gaggero:2015nsa}.
As for the CR composition, the galactic 
neutrino flux at 5 PeV could be 6 times larger if 
instead of He the dominant component beyond the CR knee
are protons, but it could also be a factor of 
0.06 smaller if they are iron nuclei. Notice that this 
uncertainty is larger than the one in 
the atmospheric flux \cite{Lipari:2013taa,Gaisser:2013ira}. 
The flux of atmospheric neutrinos is proportional to 
the all-nucleon flux, that at $E>E_{\rm knee}$ scales like
$A^{2}\Phi_A(AE)\approx A^{-1}$. In IS space, however, the situation
is quite different. First, when a nucleus collides 
most of the spectator nucleons leave unscattered keeping
all their energy; if just
one of the nucleons produced pions, 
the $\nu$ flux
would scale like  $A\Phi_A(AE)$. In addition, in IS space
the probability that the CR interacts 
grows with $A$, which provides an extra
factor of $A^{2/3}$ and a total neutrino flux proportional to 
$A^{-4/3}$. Our analysis
does not use these scaling hypothesis, it is based on EPOS-LHC
simulations of the cross sections and neutrino yields that
are specific for each
primary. The suppresion of $1/100$ that we obtain
when going from proton to Fe
is weaker than the $A^{-4/3}= 1/214$ expected from the previous scaling 
arguments, but stronger than the $A^{-1}=1/56$ present in the atmospheric
neutrino flux (see Fig.~{\ref{fig4}}).

We have found that the galactic neutrino flux is small at
all IceCube energies. More precisely, we show that at 
$E_\nu< 1$ PeV it is smaller
than the conventional flux from pion and kaon
decays, and that at 10-1000 TeV it is 
well below the expected flux from forward charm decays
(which should dominate the atmospheric neutrino flux at $E_\nu>250$ TeV). 
Of course, the angular distribution and the flavor composition
of these fluxes
do not coincide, but we estimate that among the 54 high-energy 
events observed by IceCube during the initial 4-year period, 
there could be up to 14 events \cite{Illana:2014bda}
from atmospheric neutrinos while only 1 event would have the 
galactic origin discussed here.

Although CR propagation and spallation inside our galaxy
can be simulated 
numerically with more sophisticated 
MonteCarlo codes \cite{Neronov:2013lza,Evoli:2008dv,Ahlers:2015moa}
that include local sources or leptonic gamma rays, we
think that the type of simplified analysis presented here 
may still be useful. Our results seem to agree within a 
25\% with complete GALPROP simulations, and the method that we use could
be more efficient, for example,
to identify the effects of different hypotheses
(like changes in the spectral index or in the composition of
the CR gas) or to account for the whole domain of the different
distributions involved in the calculation (notice that the 
high-energy yields extend down to $x\approx 10^{-8}$).
In this sense, we find remarkable that a change in the 
composition of an otherwise {\it smooth} CR spectrum 
can imply the sharp drop in the flux of galactic neutrinos 
at PeV energies shown in Fig.~\ref{fig5}. This is an interesting
spectral feature that could mimic
the ones expected in indirect searches for dark matter.

\acknowledgments

This work has been supported by MICINN of Spain 
(FPA2013-47836, FPA2015-68783-REDT, FPA2016-78220 and  
Consolider-Ingenio {\bf MultiDark} CSD2009-00064) and by Junta de 
Andaluc\'\i a (FQM101). JMC acknowledges a {\it Beca de
Iniciaci\'on a la Investigaci\'on} fellowship from the UGR.

\end{document}